\documentclass{iau}
\usepackage{graphicx}

\title[multiwavelength analysis of LLAGN] 
{Multiwavelength analysis as a probe of accretion and radiative
processes in LINERs}

\author[George Younes \& Delphine Porquet]   
{George Younes $^1$
 \and  Delphine Porquet $^2$}

\affiliation{$^1$ NSSTC, 320 Sparkman Drive, Huntsville, AL 35805, USA \\ email: {\tt gyounes@usra.edu } \\[\affilskip]
$^2$Observatoire Astronomique de Strasbourg, CNRS, UMR7550, 
\\ 11 rue de l'Université, 67000, Strasbourg, France \\email: {\tt delphine.porquet@astro.unistra.fr}}

\pubyear{2012}
\volume{290}  
\jname{Feeding compact objects: Accretion on all scales}
\editors{C.M. Zhang, T. Belloni, M. M\'endez \& S.N. Zhang, eds.}

\begin{document}

\maketitle

\begin{abstract}
We study the multiwavelength properties of an optically selected
sample of Low Ionization Nuclear Emission-line Regions (LINERs), in an
attempt to determine the accretion mechanism powering their central
engine. We show how their X-ray spectral characteristics, and their
spectral energy distribution compare to luminous AGN, and briefly
discuss their connection to their less massive counter-parts galactic
black-hole X-ray binaries.

\end{abstract}

\section{Introduction}

Low luminosity active galactic nuclei (LLAGN) have grown to become one
of the most interesting facets of modern AGN astrophysics. The
challenge that these fascinating objects present lies in understanding
their unusual dimness, leading to a better comprehension of the faint
end of the AGN luminosity function. Low Ionization Nuclear
Emission-line Regions (LINERs) make up 1/3 of these LLAGN and mainly
populate the nearby universe. These LINERs exhibit bolometric
luminosities at least two orders of magnitude lower than classical
luminous AGN (Seyferts and quasars). At such low luminosities, the
optically thick geometrically thin accretion disk powering luminous
AGN, is believed to switch to a geometrically thin, optically thick,
radiatively inefficient accretion flow (\cite[Narayan \& McClintock 
2008]{narayan08}). The switch occurs below  a critical mass accretion
rate $\dot{M}$, for which LINERs clearly belongs (\cite[Ho
2009]{ho09}), where the density of the disk becomes too low for
radiative cooling to be effective and most of the energy in the flow
is advected beyond the BH event horizon and/or convected away in the
form of outflows/jet (\cite[Narayan \& Yi 1995]{narayan95};
\cite[Blandford \& Begelman 1999]{blandford99}). In the following we
give the results on the X-ray and multiwavelength properties of an
optically selected sample of LINERs showing the definite detection of
broad H$\alpha$ line (LINER~1s, \cite[Ho et al. 1997]{ho97}). We
discuss these results in the context of accretion in LINERs
specifically and LLAGN in general.

\section{Results and discussion}

The X-ray spectra of all LINER~1s in our sample are fit with an
absorbed power-law, sometimes including the contribution of a thermal
component at low energies ($<$1.5~keV). They all lacked the Fe
K$\alpha$ line at 6.4 keV and the soft excess detected in most
luminous AGN. Moreover, only one of the LINER~1s showed significant
variability on time-scales $<1$ day, a popular property among Seyfert
and quasars. Finally, we found that the power-law photon index is
anticorrelated with the Eddington ratio for LINER~1s, i.e., the
spectra are hardening with increasing $L_{\rm bol}/L_{\rm Edd}$
(\cite[Younes et al. 2011]{younes11}). The opposite is found for
luminous AGN, with their X-ray spectra softening with increasing
Eddington ratio (\cite[Porquet al. 2004]{porquet04}). We note that X-ray
binaries in their low/hard state share this last similar
characteristic with LINERs (\cite[Wu \& Gu 2008]{wu08}).

The lack of an Fe line in all LINER 1s might indicate the absence of
an optically thick reflecting medium, thought to be the inner-edge of
a thin accretion disk in luminous AGN (we note, though, that only 30\%
of luminous type 1 AGN show relativistic Fe~K$\alpha$ lines,
\cite[Nandra et al. 2007]{nandra07}). The rare short time-scale
variability could be the result of a larger X-ray emitting region
compared to luminous AGN. Both properties fit the RIAF picture very
well. Additionally, the anticorrelation between $\Gamma$ and $L_{\rm
  bol}/L_{\rm Edd}$ is easily explained in the RIAF context, where the
X-ray emission is the result of inverse-Compton scattering of the
synchrotron radiation in the flow. The increase in $L_{\rm bol}/L_{\rm
  Edd}$, roughly equivalent to an increase in $\dot{M}$, will result
in more photons being scattered to higher energies by the hot electrons
in the flow, causing the X-ray spectrum to harden.

We have built the spectral energy distribution of LINER 1s in our
sample with simultaneous UV to X-ray data, mainly coming from the {\it
  XMM-Newton} telescope, eliminating the long time-scale variability
in these two bands that could bias our conclusions (\cite[Younes et
al. 2010]{younes10}; \cite[Younes et al. 2012]{younes12}). We
confirmed previous results on LINER SEDs that came from random
non-simultaneous observations; mainly, (1) at a given X-ray
luminosity, the radio emission of LINER 1s is comparable to the radio
emission of radio-loud quasars, (2) the UV big blue bump is absent with
an almost flat UV to X-ray flux ratio, $\alpha_{\rm ox}$, and (3)
their bolometric luminosity derived from their SEDs is at least two
orders of magnitude lower than luminous AGN. We found, for the first
time for LINER 1s, that $\alpha_{\rm ox}$ is anticorrelated with
Eddington ratio (\cite[Younes et al. 2012]{younes12}), in contrast to
the positive relation found for luminous AGN (\cite[Lusso et
al. 2010]{lusso10}). Such an anticorrelation for LINERs has been
predicted based on X-ray binaries in their low/hard state
(\cite[Sobolewska et al. 2011]{Sobolewska11}). Such an anticorrelation
could also be explained in the RIAF context where the X-ray luminosity
decreases faster than the UV luminosity with decreasing accretion rate
(\cite[Xu 2011]{xu11}).

\end{document}